\documentclass[showpacs,aps,pra,twocolumn,floats]{revtex4}
\usepackage{graphicx,epsfig}
\usepackage{times}
\usepackage{amsmath,amssymb,wasysym}
\usepackage{color}

\newlength\figurewidth
\newcommand{\C}{\mathbb{C}}

\newcommand{\R}{\mathbb{R}}

\newcommand{\rem}[1]{}

\newcommand{\imag}[1]{\text{Im}(#1)}

\newcommand{\real}[1]{\text{Re}(#1)}
\newcommand{\realb}[1]{\text{Re}[#1]}
\newcommand{\realc}[1]{\text{Re}\,#1}
\newcommand{\reald}[1]{\text{Re}\left[#1\right]}

\begin{document}

\title{Nonhermitian transport effects in coupled-resonator optical waveguides}
   \author{Henning Schomerus}
  \affiliation{Department of Physics, Lancaster University, Lancaster LA1 4YB, United Kingdom}
   \author{Jan Wiersig}
  \affiliation{Institut f{\"u}r Theoretische Physik, Otto-von-Guericke-Universit{\"a}t Magdeburg, Postfach 4120, D-39016 Magdeburg, Germany}
\date{\today}
\begin{abstract}
Coupled-resonator optical waveguides (CROWs) are known to have interesting and useful dispersion properties. Here, we study the transport in these waveguides in the general case where each resonator is open and asymmetric, i.e., is leaky and possesses no mirror-reflection symmetry.
Each individual resonator then exhibits asymmetric backscattering between clockwise and counterclockwise propagating waves, which in combination with the losses induces non-orthogonal eigenmodes. In a chain of such resonators, the coupling between the resonators induces an additional source of non-hermiticity, and a complex band structure arises. We show that in this situation the group velocity of wave packets differs from the velocity associated with the probability density flux, with the difference arising from a non-hermitian correction to the Hellmann-Feynman theorem.
Exploring these features numerically in a realistic scenario, we find that the complex band structure comprises almost-real branches and complex branches, which are joined by exceptional points, i.e., nonhermitian degeneracies at which not only the frequencies and decay rates coalesce but also the eigenmodes themselves.
The non-hermitian corrections to the group velocity are largest in the regions around the exceptional points.
\end{abstract}
\maketitle

\section{Introduction}
Optical microcavities attract considerable attention  due to the possibility to confine light with a well defined frequency~$\omega$ for a long time~$\tau$ to very small volumes~\cite{KV04}. These are highly desirable properties both for fundamental research and for practical device applications~\cite{Vahala03}.
One area of attention are coupled-resonator optical waveguides (CROWs), which have been proposed independently by a number of groups~\cite{LCH97,SM98,YXL99}. Such waveguides are formed by a serial chain of microcavities with high quality factors ($Q = \omega\tau$), which are weakly coupled by their evanescent fields.
Potential applications range from optical filtering~\cite{LCA04} over optical buffering~\cite{PSX04} and nonlinear components~\cite{XLY00} to group velocity compensation~\cite{KKO03}.
CROWs have been fabricated in various different cavity geometries, for instance, photonic crystal defect cavities~\cite{OSR01}, microspheres~\cite{HMT05}, microrings~\cite{PZD06}, racetrack microcavities~\cite{XSO06}, and microdisks~\cite{SCK09}.
For a recent review of the field of CROWs we refer the reader to Ref.~\cite{MFC12}.

In most studies of CROWs so far, the openness of the individual resonators has not been fully incorporated in the description. Some researchers include a decay rate of modes to account for the finite linewidths observed in transmission spectra, see, e.g., Ref.~\cite{XSO06}. On a  more fundamental level, \citet{GCR11b} investigated the impact of the decay rate on the maximum delay time achievable in CROWs.
However, the openness of a cavity not only expresses itself in a finite decay rate. Another important feature known from quasi-bound states in quantum mechanics is the appearance of nonorthogonality of modes; see, e.g., Ref.~\cite{BLP70}.
The nonorthogonality becomes extreme near so-called exceptional points (EPs) in parameter space \cite{Kato66,Heiss00,Berry04}, at which two or more eigenvalues {\it and} eigenstates coalesce.
As a matter of fact, significant nonorthogonality of nearly degenerate mode pairs appears already in slightly deformed or perturbed microdisk cavities which do not possess a mirror-reflection symmetry~\cite{WKH08,WES11,Wiersig11}. This interesting phenomenon has been traced back to {\it asymmetric backscattering} of clockwise (CW) and counterclockwise (CCW) propagating traveling waves~\cite{Wiersig14}.
A consequence of this asymmetric backscattering is that both modes have a similar finite orbital angular momentum, meaning that both modes mainly copropagate in the same direction. This property is known as ``chirality'', but should not to be confused with the optical activity in chiral media, see, e.g., Ref.~\cite{Lekner96}.  Near an EP, the above-mentioned chirality can be related to the intrinsic chirality defined in the space of the near-degenerate modes \cite{HH01,DDG03}.
Chirality of modes has been confirmed recently in an experiment on a microdisk with two Rayleigh scatterers on the perimeter~\cite{KKS14}, while EPs in complex band structures have been observed in numerical simulations of gain-loss modulated optical lattices~\cite{MEC08,MEC10,BHS10,SRB11,RKK12}. However, up to now these nonorthogonality effects have been ignored in the modeling of CROWs.

\begin{figure}[t]
\includegraphics[width=\figurewidth]{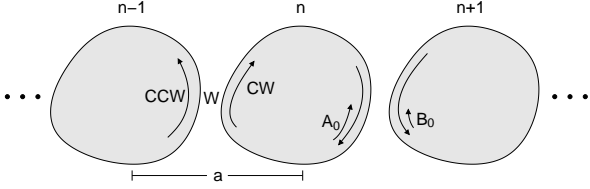}
\caption{Sketch of a coupled-resonator optical waveguide with asymmetric resonators. The index $n$ labels the unit cell of size $a$. The parameter  $A_0$ describes the internal backscattering of clockwise (CW) traveling waves to counterclockwise (CCW) traveling waves, while the backscattering from counterclockwise to clockwise traveling waves is described by $B_0$. Furthermore, $W$ is the inter-cavity coupling strength. }
\label{fig:chain}
\end{figure}

The aim of this paper is to provide a general description of wave propagation along CROWs in the presence of  non-hermitian effects, which may arise both within each resonator as well as from the coupling between the resonators.
First, we formulate a general theory of wave propagation in a CROW  tight-binding chain with a complex band structure and non-orthogonal modes. We show that these properties give rise to systematic corrections to the group velocity, which no longer coincides with the propagation velocity inferred from the probability flux. These corrections find a natural interpretation in a non-hermitian reformulation of the Hellmann-Feynman theorem.
We then describe how the desired non-hermitian effects arise generically
in an arrangement of a CROW with asymmetric coupled microresonators, as illustrated in Fig.~\ref{fig:chain}.
Such systems combine losses, asymmetric internal backscattering within each resonator, and asymmetric coupling between the resonators, and thus display EPs both for the individual resonators as well as for the coupled chain.
These findings are illustrated numerically for the specific implementation of a CROW formed by microdisks that are perturbed by nanoparticles. We show that the non-hermitian effects combine quite naturally to induce EPs in the complex band structure, which displays almost-real branches and complex branches.
The non-hermitian corrections to the group velocity are largest in the regions around the EPs, and still substantial in the complex branches of the dispersion.

The article is organized as follows.
In Sec.~\ref{sec:theory} we formulate our theory of wave propagation in CROWs, where we account for non-hermitian effects both within a resonator as well as for the coupling between the resonators.
Section~\ref{sec:CROW}  describes how these non-hermitian effects arise in CROWs with asymmetric internal backscattering.
Section~\ref{sec:disk} presents the numerical results for band structure and wave propagation for the CROW formed by microdisks perturbed by nanoparticles. Concluding remarks are given in Sec.~\ref{sec:discussion}.

\section{General theory of transport in nonhermitian resonator chains}
\label{sec:theory}
\subsection{Model}
We consider a multi-mode wave $|\psi\rangle=\sum_n\boldsymbol\psi_n|n\rangle$   that propagates along a chain of sites with index $n$.
In each unit cell the wave has $M$ components, collected into a vector $\boldsymbol\psi_n$.
The wave propagation is governed by the evolution equation
\begin{eqnarray}\label{eq:weq}
|\dot\psi\rangle&=&-i[\hat C|\psi\rangle+ \hat T^{(+)}|\psi^{(+)}\rangle+ \hat T^{(-)}|\psi^{(-)}\rangle],
\end{eqnarray}
where the wave functions $|\psi^{(\pm)}\rangle=\sum_n\boldsymbol\psi_{n\pm1}|n\rangle$ are obtained from $|\psi\rangle$ via a shift by one site to the right or left.
The operators $\hat C$ and $\hat T^{(\pm)}$ act in the mode space on each lattice site, where they are associated to $M\times M$ matrices $C$ and $T^{(\pm)}$.
Therefore, $\hat C$ describes the coupling of modes in a given unit cell while $\hat T^{(\pm)}$ describe the coupling to the neighboring unit cells.
In the hermitian situation, $\hat C=\hat C^\dagger$ and $\hat T^{(-)}=\hat T^{(+)\dagger}$. We are, however, interested in the general case where these relations do not hold, and aim to characterize the wave propagation by the evolution of the position expectation value, which we  contrast  with the probability flux that appears in the continuity equation.


\subsection{Nonhermitian Ehrenfest theorem for the position expectation value}
The position operator $\hat x=a \hat n=a\sum_n n|n\rangle \langle n|$ acts in lattice-site space, where $n$ is the site index and $a\equiv 1$ is the lattice spacing.
The expectation value of position is defined as
\begin{equation}\label{eq:x}
\langle x\rangle= \frac{\langle \psi | \hat n|\psi\rangle}{\langle \psi|\psi\rangle}.
\end{equation}
With Eq. \eqref{eq:weq} the derivative is then given by
\begin{eqnarray}
&&\frac{d}{dt}\langle x\rangle
\nonumber\\
&&= i \frac{[\langle \psi|\hat C^\dagger\hat n|\psi\rangle  +\langle \psi^{(+)}|\hat T^{(+)\dagger}\hat n|\psi\rangle +\langle \psi^{(-)}|\hat T^{(-)\dagger}\hat n|\psi\rangle ]}{\langle \psi|\psi\rangle}
\nonumber
\\&&
-i \frac{[\langle \psi| \hat n\hat C |\psi\rangle+ \langle \psi| \hat n\hat T^{(+)}|\psi^{(+)}\rangle+\langle \psi| \hat n\hat T^{(-)}|\psi^{(-)}\rangle]}{\langle \psi|\psi\rangle}
\nonumber
\\&&
-  i \frac{[\langle \psi|\hat C^\dagger|\psi\rangle  +\langle \psi^{(+)}|\hat T^{(+)\dagger}|\psi\rangle +\langle \psi^{(-)}|\hat T^{(-)\dagger}|\psi\rangle]\langle \psi|\hat n|\psi\rangle}{\langle \psi|\psi\rangle^2}
\nonumber
\\&&
+  i \frac{[\langle \psi| \hat C |\psi\rangle+ \langle \psi| \hat T^{(+)}|\psi^{(+)}\rangle+\langle \psi| \hat T^{(-)}|\psi^{(-)}\rangle]\langle \psi|\hat n|\psi\rangle}{\langle \psi|\psi\rangle^2}.
\nonumber
\\
\label{eq:x2}
\end{eqnarray}
The last two lines arise due to the time dependence of the wave function normalization in the denominator of Eq.~\eqref{eq:x}.

\subsection{Evolution of a wave packet}
We now evaluate the terms in Eq.\ \eqref{eq:x2} for a wave packet of the form
\begin{eqnarray}
\boldsymbol\psi_n(t)&=& c\int_{BZ} dk' \mathbf{v}(k') \phi(k') e^{ik'n},\\
 \phi(k')&=&\exp[-i\varepsilon(k')t-ik'x-\sigma^2(k'-k)^2],
\end{eqnarray}
which is centered around position $x$ and wave number $k$ and has a width $\sigma$ in lattice space.
Here $\varepsilon(k)$ is an eigenvalue of the Bloch Hamiltonian
\begin{equation}\label{eq:bloch2}
H(k)=C+T^{(+)}e^{ik}+T^{(-)}e^{-ik}
\end{equation}
and $\mathbf{v}(k)$ the corresponding normalized eigenvector ($\mathbf{v}(k)^\dagger \mathbf{v}(k) = 1$). Each eigenvalue defines a different wave packet, associated to a different branch (band) of the dispersion relation. In the nonhermitian case, the eigenvalues are in general complex, and the associated eigenvectors are no longer mutually orthogonal to each other.

We are interested in the limit $\sigma\to\infty$, resulting in a wave packet with a well defined wave number. For any operator $\hat M$ which solely acts in mode space and there is associated with a matrix $M$, we then obtain
\begin{align}
&\langle \psi|\hat M|\psi\rangle
\nonumber \\&=\int dk''dk' \mathbf{v}(k'')^\dagger M \mathbf{v}(k') \phi^*(k'')\phi(k')  \sum_n e^{i(k'-k'')n}
\nonumber \\
&= \mathbf{v}^\dagger M \mathbf{v},
\end{align}
where we used  the Poisson summation formula $\sum_n  e^{i(k'-k'')n}=2\pi \sum_m\delta(k'-k''-2\pi m)$  and denoted
$\mathbf{v}=\mathbf{v}(k)$. Here and henceforth we fixed $c$ to ensure normalization for the choice $M=\openone$ at the given time $t$, thereby removing the factor $\exp[2{\rm Im}\,\varepsilon(k)t]$ encountered for a complex dispersion relation. (The normalization constant $c$ drops out of the ratios in the expectation values and thus can be chosen in this convenient time-dependent form.)

Furthermore, using $\sum_n in e^{i(k'-k'')n}=\sum_n \frac{1}{2}(\partial_{k'}-\partial_{k''}) e^{i(k'-k'')n}$ and integrating by parts we can write
\begin{eqnarray}
&&i\langle \psi|\hat M\hat n|\psi\rangle\nonumber \\
&&=\int dk''dk' \mathbf{v}(k'')^\dagger M \mathbf{v}(k') \phi^*(k'')\phi(k')  \sum_n in e^{i(k'-k'')n} \nonumber \\
&&= \pi \int dk (\partial_{k''}-\partial_{k'})[\mathbf{v}(k'')^\dagger M \mathbf{v}(k') \phi^*(k'')\phi(k')]|_{k'=k''=k},\nonumber \\
\end{eqnarray}
which in the limit $\sigma\to \infty$ approaches
\begin{eqnarray}
&&i\langle \psi|\hat M\hat n|\psi\rangle
=-\frac{1}{2}(\partial_{k'}-\partial_{k''})[\mathbf{v}(k'')^\dagger M \mathbf{v}(k')]|_{k'=k''=k}
\nonumber \\
&&-2 \pi \mathbf{v}^\dagger M \mathbf{v} \int dk \frac{1}{2}(\partial_{k'}-\partial_{k''})[ \phi^*(k'')\phi(k')]|_{k'=k''=k}
\nonumber\\&&=
\frac{1}{2}[\mathbf{v}^{\prime\dagger} M \mathbf{v}-\mathbf{v}^{\dagger} M \mathbf{v}']
-\mathbf{v}^\dagger M \mathbf{v} I
.\end{eqnarray}
Here we again used the value of $c$ implied by normalization and denoted $d\mathbf{v}/dk =\mathbf{v}'$.
The integral $I$ will drop out of the final expressions and thus does not need to be evaluated explicitly.

An analogous calculation can be carried out for the terms containing $\psi^{(+)}$. In $k$-space, $\psi^{(+)}$ contributes an additional factor $e^{ik}$, which can be transferred to the scalar products by attributing it to $M$.
Therefore,
\begin{eqnarray}
 &&
 \langle \psi| \hat M |\psi^{(\pm)}\rangle
= e^{\pm ik}\mathbf{v}^\dagger M \mathbf{v}
\end{eqnarray}
while
\begin{align}
& i\langle \psi| \hat M \hat n|\psi^{(\pm)}\rangle
\nonumber \\
&=
\frac{1}{2}[\mathbf{v}^{\prime\dagger} M \mathbf{v}-\mathbf{v}^{\dagger} M \mathbf{v}']e^{\pm ik}\mp\frac{i}{2}\mathbf{v}^\dagger M \mathbf{v} e^{\pm ik}
-\mathbf{v}^\dagger M \mathbf{v} e^{\pm ik} I
.\nonumber \\
\end{align}
It is useful to note that this can be written as
\begin{equation}
 i\langle \psi| \hat M \hat n|\psi^{(\pm)}\rangle
=i\langle \psi|  \hat M \hat n|\psi\rangle e^{\pm ik}\mp\frac{i}{2}\langle \psi|  \hat M |\psi\rangle e^{\pm ik}.
\end{equation}

In Eq.~\eqref{eq:x2}, these terms then combine to our preliminary result
\begin{align}
&\frac{d}{dt}\langle x\rangle= i\langle\psi|(\hat H^\dagger- \hat H)\hat n|\psi\rangle  +\frac{1}{2}\langle \psi|\frac{d\hat  H}{dk}+\frac{d \hat H^\dagger}{dk}|\psi\rangle
\nonumber\\
&\quad{}- i\langle\psi|(\hat  H^\dagger-\hat  H)|\psi\rangle\langle\psi| \hat n|\psi\rangle
\\
&=\frac{1}{2}
\mathbf{v}^{\prime\dagger}(H^\dagger-H)\mathbf{v} - \frac{1}{2}
\mathbf{v}^{\dagger} (H^\dagger-H)\mathbf{v}'
\nonumber\\
&
\quad{}+\frac{1}{2}\mathbf{v}^\dagger(\frac{d\hat  H}{dk}+\frac{d \hat H^\dagger}{dk})\mathbf{v}
- \mathbf{v}^\dagger (H^\dagger-H)\mathbf{v}\frac{1}{2}(\mathbf{v}^{\prime\dagger}\mathbf{v}-\mathbf{v}^\dagger \mathbf{v}')
\\
&=
\,{\rm Re}\, [\mathbf{v}^\dagger  \frac{dH}{dk} \mathbf{v}]
+{\rm Re}\,[\mathbf{v}^{\dagger} (H-H^\dagger)  \mathbf{v}']
+2\,{\rm Re}\,[(\mathbf{v}^\dagger H \mathbf{v}) (\mathbf{v}^{\prime\dagger}\mathbf{v})].
\label{eq:vprefinal}
\end{align}
In the last step we used $\mathbf{v}^{\prime\dagger} \mathbf{v}=-\mathbf{v}^\dagger \mathbf{v}'$ for normalized $\mathbf{v}$.

\subsection{Interpretation as a generalized group velocity}

We now can show that the final expression equates to the suitably generalized group velocity
\begin{equation}\label{eq:defvg}
v_g\equiv\frac{d}{dk}{\rm Re}\,\varepsilon(k)
\end{equation}
 associated to the eigenvalue $\varepsilon(k)$. Writing
\begin{equation}
{\rm Re}\,\varepsilon(k)={\rm Re}\,\mathbf{v}^\dagger H \mathbf{v}
\end{equation}
as implied by the eigenvalue condition $H \mathbf{v}=\varepsilon \mathbf{v}$, we find
\begin{align}
v_g& ={\rm Re}\,\mathbf{v}^{\dagger}(dH/dk)\mathbf{v}+{\rm Re}\,\mathbf{v}^{\prime\dagger}H\mathbf{v}+
{\rm Re}\,\mathbf{v}^{\dagger}H \mathbf{v}'
\\& ={\rm Re}\,[\mathbf{v}^{\dagger}(dH/dk)\mathbf{v}]+{\rm Re}\,[\mathbf{v}^{\dagger}(H-H^\dagger)\mathbf{v}']
+2\,{\rm Re}\,[\mathbf{v}^{\prime\dagger}H\mathbf{v}],
\end{align}
which coincides with Eq.\ \eqref{eq:vprefinal} as the eigenvalue condition implies $\mathbf{v}^{\prime\dagger}H\mathbf{v}=\varepsilon(k) \mathbf{v}^{\prime\dagger}\mathbf{v} =(\mathbf{v}^\dagger H \mathbf{v}) (\mathbf{v}^{\prime\dagger}\mathbf{v})$.
Finally, as ${\rm Re}\,[\mathbf{v}^{\prime\dagger}H\mathbf{v}]={\rm Re}\,[\mathbf{v}^{\dagger}H^\dagger \mathbf{v}']$  the group velocity can be rewritten as
\begin{eqnarray}\label{eq:vfinal}
v_g={\rm Re}\,[\mathbf{v}^{\dagger}(dH/dk)\mathbf{v}]+{\rm Re}\,[\mathbf{v}^{\dagger}(H+H^\dagger)\mathbf{v}'].
\end{eqnarray}
The first term corresponds to the Hellmann-Feynman theorem, while the second term is a correction arising due to the complex eigenvalues and nonorthogonal eigenstates in the nonhermitian case. While various generalizations of the Hellmann-Feynman theorem in different contexts are known~\cite{ZKM87,EFC10,GS11}, our compact results are specific to the nonhermitian dynamics of a wave packet with a well-defined wave number.

\subsection{Comparison to the probability flux}

It is  instructive  to compare the wave-packet propagation velocity $v_g$ to the probability flux associated to the density $\rho_n=\boldsymbol\psi_n^\dagger\boldsymbol\psi_n$. From Eq.~\eqref{eq:weq} we find the continuity equation
\begin{align}\label{eq:conteq}
\dot \rho_n&=&i\boldsymbol\psi_n^\dagger(C^\dagger-C)\boldsymbol\psi_n-(J_{n+1/2}^{(+)}-J_{n-1/2}^{(-)}).
\end{align}
The first term is a source term generated within the unit cell, and vanishes  in the hermitian case. The second and third terms are the probability fluxes from the left and right neighbor cell, which we denote as
\begin{align}\label{eq:Jplus}
J_{n+1/2}^{(+)}&=&i(\boldsymbol\psi_n^\dagger T^{(+)}\boldsymbol\psi_{n+1}-\boldsymbol\psi_{n+1}^\dagger T^{(+)\dagger}\boldsymbol\psi_n),\\
\label{eq:Jminus}
J_{n+1/2}^{(-)}&=&i(\boldsymbol\psi_{n}^\dagger T^{(-)\dagger}\boldsymbol\psi_{n+1}-\boldsymbol\psi_{n+1}^\dagger T^{(-)}\boldsymbol\psi_{n}).
\end{align}

In the nonhermitian case, these two terms are fully independent of each other, i.e., the flux from cell $n$ to $n\pm 1$ does not equate to the flux from cells  $n\pm 1$ to cell $n$.

In order to identify the flux contribution in the propagation velocity, we write
expression \eqref{eq:x2} in the form
\begin{widetext}
\begin{align}
\label{eq:symversion}
\frac{d}{dt}\langle x\rangle=&
\frac{i}{2} \frac{\langle \psi^{(-)}|\hat T^{(-)\dagger} |\psi \rangle  -\langle\psi| \hat T^{(-)}|\psi^{(-)} \rangle+\langle \psi|\hat T^{(+)} |\psi^{(+)} \rangle  -\langle\psi^{(+)}| \hat T^{(+)\dagger}|\psi \rangle}{\langle \psi|\psi\rangle}
\nonumber
\\&+ \frac{i}{2} \frac{[2\langle\psi|\hat D\hat n|\psi \rangle +\langle \psi^{(+)}|\hat V \hat n|\psi \rangle-\langle\psi|\hat V^\dagger \hat n|\psi^{(+)} \rangle+\langle \psi|\hat V \hat n|\psi^{(-)} \rangle-\langle\psi^{(-)}|\hat V^\dagger \hat n|\psi \rangle]}{\langle \psi|\psi\rangle}
\nonumber
\\&
- \frac{i}{2}\frac{[2\langle\psi|\hat D|\psi \rangle +\langle \psi^{(+)}|\hat V |\psi \rangle-\langle\psi|\hat V^\dagger |\psi^{(+)} \rangle+\langle \psi|\hat V |\psi^{(-)} \rangle-\langle\psi^{(-)}|\hat V^\dagger |\psi \rangle]\langle \psi|\hat n|\psi\rangle  }{\langle \psi|\psi\rangle^2}
\end{align}
\end{widetext}
where $\hat D=\hat C^\dagger-\hat C$ and $\hat V=\hat T^{(+)\dagger}- \hat T^{(-)}$.
The first line of terms involves the probability flux terms defined above. The second line is a contribution to the dynamics of the numerator in \eqref{eq:x}, while the third line arises due to the time dependence of the normalization factor in the denominator.
In the hermitian case, $\hat D$ and $\hat V$ vanish, so that the group velocity equates to the flux contribution, $v_I$.
In the general nonhermitian case, considering the wave packet defined above
\begin{align}
v_I=&
\frac{i}{2} \mathbf{v}^\dagger(   T^{(-)\dagger}  e^{ik} -T^{(-)}  e^{-ik} +T^{(+)}  e^{ik} -  T^{(+)\dagger} e^{-ik})\mathbf{v}
\\=&\label{eq:vI}
{\rm Re}\,[\mathbf{v}^\dagger (dH/dk) \mathbf{v}],
\end{align}
as in the original Hellmann-Feynman theorem.
Comparison with Eq.~\eqref{eq:vfinal} gives the difference
\begin{eqnarray}\label{eq:vdiff}
v_g-v_I={\rm Re}\,[\mathbf{v}^{\dagger}(H+H^\dagger)\mathbf{v}'],
\end{eqnarray}
which corresponds to the nonhermitian correction  to the Hellmann-Feynman theorem.

\subsection{Behaviour near exceptional points}

The correction \eqref{eq:vdiff} originates in the non-orthogonality of the eigenvectors for the non-hermitian Bloch Hamiltonian \eqref{eq:bloch2}, and thus can only be nonzero for multi-mode waves, i.e., $M>1$. The non-orthogonality becomes most pronounced
near EPs, which are the generic degeneracy points of non-hermitian systems. At an EP of order $p\leq M$, not only $p$ eigenvalues coalesce, but also the eigenvectors become degenerate \cite{Kato66,Heiss08},
\begin{eqnarray}
\varepsilon(k)-\varepsilon_{\text{EP}} & \sim &  (k-k_{\text{EP}})^{1/p} \ ,\\
\mathbf{v}(k)-\mathbf{v}_{\text{EP}}  & \sim & (k-k_{\text{EP}})^{1/p} \ .
\end{eqnarray}
The group velocity follows then from Eq.~(\ref{eq:defvg})
\begin{equation}\label{eq:vgEP}
v_g(k) \sim (k-k_{\text{EP}})^{(1/p-1)}
\end{equation}
which diverges when the EP is approached. It should be emphasised that this divergence is not related to a divergence of $dH/dk$, which remains finite. Therefore the intensity transport velocity (Eq.~(\ref{eq:vI})) near an EP
\begin{equation}\label{eq:vIEP}
v_I(k) = v_{I, \text{EP}} + \beta(k-k_{\text{EP}})^{1/p}
\end{equation}
remains finite as well, with some constants $v_{I, \text{EP}}$ and $\beta$. This implies that the correction term in Eq.~(\ref{eq:vdiff}) diverges at the EPs of the Bloch Hamiltonian, which emphasizes the role of mode non-orthogonality.

\section{Adaptation to asymmetric coupled-resonator optical waveguides}
\label{sec:CROW}
We now describe how the nonhermitian features entering the general model arise in the context of coupled-resonator optical waveguides with asymmetric internal backscattering, constituted by coupled two-dimensional microdisks  as sketched in Fig.~\ref{fig:chain}.
This arrangement allows to naturally break all symmetries, including symmetries in an enlarged unit cell.
(A counter example would be CROW fabricated microspirals with double notches~\cite{LP09}, which possess a mirror-reflection symmetry in an enlarged unit cell.)

\subsection{Two-mode model for isolated cavity}
We first  briefly review the two-mode model for an individual fully asymmetric two-dimensional microdisk as developed in Refs.~\cite{WKH08,WES11,Wiersig11,Wiersig14}.
In this model the effective index approximation is employed. Here, the solutions of Maxwell's equations with harmonic time dependence $e^{-i\omega t}$ -- the optical modes -- can be expressed by a complex-valued wave function $\psi$. In the case of transverse magnetic (TM) polarization $\psi$ determines the electric field vector $\vec{E}(x,y,t) \propto (0,0,\realb{\psi(x,y)e^{-i\omega t}})$ perpendicular to the cavity plane. For  transverse electric (TE) polarization, $\psi$ determines the magnetic field vector $\vec{H}(x,y,t) \propto (0,0,\realb{\psi(x,y)e^{-i\omega t}})$.
In the following we use the dimensionless frequency $\Omega = \omega R/c$ with $c$ being the speed of light in vacuum and $R$ being a length scale of the problem, e.g., the radius in the case of a circular disk.

In the slowly-varying envelope approximation in the time domain~\cite{Siegman86}, the dynamics of the wave function~$\psi$  is described by a Schr\"odinger-type equation
\begin{equation}\label{eq:slowlyvarying}
i\frac{d}{dt}\psi(x,y) = H\psi(x,y) \ .
\end{equation}
Starting from Maxwell's equations, Eq.~(\ref{eq:slowlyvarying}) is derived by assuming that the optical field varies slowly in time (not necessarily in space) with respect to a reference frequency.

In the two-mode approximation, the Hamiltonian for the isolated cavity is in the CCW/CW traveling-wave basis
\begin{equation}\label{eq:isolatedcavity}
H_1 = \left(\begin{array}{cc}
\Omega_0 & A_0     \\
B_0      & \Omega_0\\
\end{array}\right) \ .
\end{equation}
The dimensionless diagonal element $\Omega_0\in\C$ describes the frequency of the CCW and CW components in the absence of backscattering. The backscattering is described by the dimensionless off-diagonal elements $A_0, B_0 \in\C$. In general, the backscattering is asymmetric with $|A_0| \neq |B_0|$.

The eigenvalues and eigenvectors of the Hamiltonian in Eq.~(\ref{eq:isolatedcavity}) are given by
\begin{equation}\label{eq:eigenvalues}
\Omega_\pm = \Omega_0\pm\sqrt{A_0B_0} \ ,
\end{equation}
\begin{equation}\label{eq:eigenvectors}
 \mathbf{v}_\pm = \eta_0\left(\begin{array}{c}
\sqrt{A_0}\\
\pm\sqrt{B_0}\\
\end{array}\right) \ ,
\end{equation}
where $\eta_0=(|A_0|+|B_0|)^{-1/2}$ is the normalization constant.
If $|A_0| \neq |B_0|$ there is an imbalance of CCW (intensity $|A_0|$) and CW (intensity $|B_0|$) components. According to Ref.~\cite{KKS14} we quantify this imbalance by the chirality
\begin{equation}\label{eq:localchirality}
\alpha = \frac{|A_0|-|B_0|}{|A_0|+|B_0|} \ .
\end{equation}
In contrast to the original definition of the chirality \cite{WKH08,WES11,Wiersig11,Wiersig14}, this chirality provides information on the sense of rotation. In the case where the CCW (CW) component dominates, $|A_0| \gg |B_0|$ ($|A_0| \ll |B_0|$), the chirality approaches $1$ ($-1$). For a balanced contribution, $|A_0| \approx |B_0|$, the chirality is close to 0. Note that both modes show the same chirality, which means in particular that their main propagation direction is the same.

The scalar product between the two normalized eigenvectors
\begin{equation}\label{eq:overlap}
S = | \mathbf{v}^\dagger_+\cdot \mathbf{v}_-|
\end{equation}
is related to the chirality by $S = |\alpha|$. That implies that when CW and CCW components are imbalanced then the mode pair is significantly nonorthogonal. When the limit of perfect chirality $|\alpha| \to 1$ is approached the modes become collinear, $S\to 1$. In this limit we therefore approach an EP with order $p=2$ where eigenvalues and eigenvectors coalesce, see Eqs.~(\ref{eq:eigenvalues})-(\ref{eq:eigenvectors}).

\subsection{Tight-binding model for CROW}
Next we discuss the coupling of asymmetric cavities in the tight-binding approximation. We start with the simplest case, two cavities. We assume that the coupling region of the cavities is sufficiently long compared to the wavelength~$\lambda$, so that the light traveling CCW (CW) in a cavity couples only to the CW (CCW) traveling wave in the adjacent cavity, cf. Fig.~\ref{fig:chain}. The coupling matrix of two neighboring cavities therefore is
\begin{equation}\label{eq:twocavitycouplingmatrix}
H_2 = \left(\begin{array}{cccc}
\Omega & A_0    & 0      & W     \\
B_0    & \Omega & W      & 0     \\
0      & W      & \Omega & A_0   \\
W      & 0      & B_0    & \Omega\\
\end{array}\right)
\end{equation}
with inter-cavity coupling coefficient $W\in\C$. Usually, $|\imag{W}| \ll |\real{W}|$, while the case of nonzero $\imag{W}$ is linked to an effect called $Q$-splitting~\cite{Wiersig06,BSW11}. Note that $\Omega_0$ is replaced by $\Omega$ to allow for a modification of the diagonal elements induced by the coupling.

For the case of the infinite chain we write Eq.~(\ref{eq:weq}) as
\begin{eqnarray}
\label{eq:chain1}
i\dot{a}_n & = & \Omega a_n + A_0b_n + Wb_{n+1}+Wb_{n-1}\\
\label{eq:chain2}
i\dot{b}_n & = & \Omega b_n + B_0a_n + Wa_{n+1}+Wa_{n-1}
\end{eqnarray}
where $a_n$ ($b_n$) is the CCW (CW) component in the $n$th cavity.
This corresponds to the tight-binding chain \eqref{eq:weq}, with $\boldsymbol\psi_n=(a_n,b_n)^T$ a two-component vector coupled by matrices
\begin{equation}
C=\left(\begin{array}{cc}
\Omega & A_0     \\
B_0      & \Omega\\
\end{array}\right)
\end{equation}
and
\begin{equation}
T^{(+)}=T^{(-)}=  \left(\begin{array}{cc}
0      & W     \\
 W      & 0 \\
\end{array}\right).
\end{equation}
Note that in general $C\neq C^\dagger$, while $T^{(-)}= T^{(+)\dagger}$ only if $W$ is real.

The density is expressed as
\begin{equation}\label{eq:density}
\rho_n = |a_n|^2+|b_n|^2 \ .
\end{equation}
As in Eq.~(\ref{eq:conteq}) we calculate the discrete continuity equation
with the current densities (cf. Eqs.~(\ref{eq:Jplus}) and (\ref{eq:Jminus}))
\begin{eqnarray}\label{eq:currentplus}
J^{(+)}_{n+1/2} & = & -2\,\imag{Wa_n^*b_{n+1}+Wb_n^*a_{n+1}} \\
\label{eq:currentminus}
J^{(-)}_{n+1/2} & = & -2\,\imag{W^*a_n^*b_{n+1}+W^*b_n^*a_{n+1}} \ .
\end{eqnarray}
In the special case $W\in\R$ we obtain $J^{(+)}_{n+1/2} = J^{(-)}_{n+1/2}$.

The Bloch-mode solutions of Eqs.~(\ref{eq:chain1})-(\ref{eq:chain2}) are of the type
\begin{equation}
\label{eq:bloch}
a_n = a(k)e^{ink}
\;\;,\;
b_n = b(k)e^{ink} \ .
\end{equation}
Inserting these Bloch modes into Eqs.~(\ref{eq:chain1})-(\ref{eq:chain2}) gives the differential equation
\begin{equation}\label{eq:ode}
i\frac{d}{dt}\left(\begin{array}{c}
a(k)\\
b(k)\\
\end{array}\right)
= H(k)
\left(\begin{array}{c}
a(k)\\
b(k)\\
\end{array}\right)
\end{equation}
with the $k$-dependent Bloch Hamiltonian
\begin{equation}\label{eq:Hamk}
H(k)
= \left(\begin{array}{cc}
\Omega       & A(k) \\
B(k) & \Omega       \\
\end{array}\right)
\end{equation}
and
\begin{eqnarray}
A(k) & = & A_0 + 2W\cos{k}, \\
B(k) & = & B_0 + 2W\cos{k} \ .
\end{eqnarray}
Note that the Bloch Hamiltonian~(\ref{eq:Hamk}) has the same structure as the Hamiltonian of the isolated cavity in Eq.~(\ref{eq:isolatedcavity}).

For the solutions of Eq.~(\ref{eq:ode}) with harmonic time dependence we compute the eigenvalues of $H(k)$
\begin{equation}\label{eq:eigenvaluesolution}
\Omega_\pm(k) = \Omega \pm \sqrt{A(k)B(k)} \ .
\end{equation}
In follows that the band structure is symmetric around the center of the Brillouin zone $k=0$, which is nontrivial considering that the individual cavities are asymmetric.

The eigenvectors of the Hamiltonian~(\ref{eq:Hamk}) are
\begin{equation}\label{eq:eigenvector}
\mathbf{v}_\pm(k)=\left(\begin{array}{c}
a_\pm(k) \\
b_\pm(k) \\
\end{array}\right)
= \eta(k)\left(\begin{array}{c} \sqrt{A(k)} \\ \pm\sqrt{B(k)}\end{array}\right) \ ,
\end{equation}
where $\eta(k)=(|A(k)|+|B(k)|)^{-1/2}$ is the normalization constant. The $k$-dependent chirality follows as
\begin{equation}\label{eq:chirality}
\alpha(k) = \frac{|A(k)|-|B(k)|}{|A(k)|+|B(k)|} \ .
\end{equation}
Note that according to the Bloch-structure in Eq.~(\ref{eq:bloch}), the chirality of waves in each cavity equals the $k$-dependent chirality.

From expression \eqref{eq:vI} in our general theory, the transport velocity of intensity follows as
\begin{equation}\label{eq:current_bloch}
v_{I\pm}(k) = \mp 4\,\frac{\reald{W}\reald{\sqrt{A^*(k)B(k)}}}{|A(k)|+|B(k)|}\sin{k} \ .
\end{equation}
This expression agrees with the averaged flux from Eqs.\ \eqref{eq:currentplus}, \eqref{eq:currentminus}. A straightforward calculation shows that the intensity transport velocity remains bounded,
\begin{equation}\label{eq:vmax}
|v_{I\pm}(k)| \leq 2|W| \ .
\end{equation}
The group velocity \eqref{eq:vfinal} is
\begin{eqnarray}\label{eq:group_velocity}
v_{g\pm}(k)
 & = & \mp \reald{W\frac{A(k)+B(k)}{\sqrt{A(k)B(k)}}}\sin{k} \ ,
\end{eqnarray}
which coincides with a direct calculation from $v_{g\pm}(k)=\frac{d\real{\Omega_\pm(k)}}{dk}.$

The difference between both velocities can be written as
\begin{eqnarray}\label{eq:vdiff2}
&&v_{g\pm}(k)-v_{I\pm}(k) =  \mp\frac{\alpha(k)\sin{k}}{|A(k)B(k)|}\\
&& \times\reald{\sqrt{A^*(k)B(k)}(W^*|A(k)|-W|B(k)|)},\nonumber
\end{eqnarray}
which after some algebra is seen to be in agreement with the general expression \eqref{eq:vdiff}.

Note that the difference between the two velocities disappears at points in the dispersion where the $k$-dependent chirality $\alpha(k)$ vanishes, i.e.,
$|A(k)|=|B(k)|$. This typically occurs at isolated points, even in the case of symmetric backscattering in each isolated resonator ($|A_0|=|B_0|$ for the $k$-independent single-resonator parameters). The two velocities only agree across the whole dispersion if $A_0=B_0$ for the isolated resonator, meaning $A(k)=B(k)$. This does not require hermiticity as $A_0=B_0$ may still have a complex phase, leading to a fully complex dispersion relation.

Another instructive limiting case is $|A_0|, |B_0| \ll 2|W|$ and $\cos{k} \approx \pm 1$, i.e., in the center and at the border of the Brillouin zone. There we find
\begin{equation}\label{eq:comparison_velocities}
v_{I\pm}(k) = v_{g\pm}(k) = \mp 2\realc{W}\sin{k}\ .
\end{equation}

\begin{figure}[b]
\includegraphics[width=0.6\figurewidth]{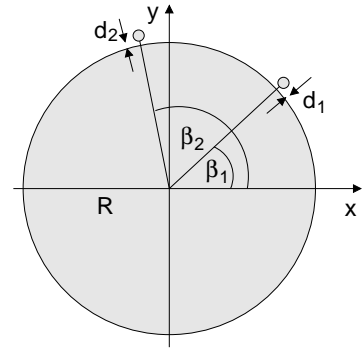}
\caption{Microdisk of refractive index $n$ and radius $R$ with two nanoparticles of refractive index $n_j$ and radii $r_j$ at distance $d_j$ from the microdisk. The azimuthal position of the nanoparticles is specified by the angles~$\beta_j$.}
\label{fig:sketch}
\end{figure}

\section{Implementation for microdisks perturbed by nanoparticles}
\label{sec:disk}
\subsection{Determination of tight-binding parameters}
Here we consider a particular unit cell, a microdisk perturbed by two nanoparticles; see Fig.~\ref{fig:sketch}. The purpose is to determine the tight-binding parameters and to demonstrate the validity of the two-cavity coupling matrix~(\ref{eq:twocavitycouplingmatrix}). We follow Ref.~\cite{Wiersig11}, where  this program was carried out for the case of a  single cavity of this particular geometry, described by Eq.~(\ref{eq:isolatedcavity}). The idea is the following. For the microdisk without nanoparticles we can choose the given mode pair with frequency $\omega_0$ and azimuthal mode number $m$ such that we have standing-wave modes with even and odd parity with respect to the horizontal $x$-axis. The even parity mode has a $\cos{m\phi}$ dependence and the odd parity mode a $\sin{m\phi}$ dependence. Placing a single nanoparticle with radius $r_1$ somewhere on the line $x=0$ does not couple even and odd parity modes provided that the nanoparticle itself has the reflection symmetry, which we can always assume since in the limit of Rayleigh scattering ($r_1 \ll \lambda$) the shape of the nanoparticle does not matter.  The perturbation projected onto the two-dimensional standing-wave basis [even mode: $(1,0)$, odd mode: $(0,1)$] can therefore be written as
\begin{equation}\label{eq:H1}
\tilde{h}_1 = \left(\begin{array}{cc}
2V_1 & 0 \\
0  & 2U_1\\
\end{array}\right)
\end{equation}
with $U_1, V_1\in\C$, where $|U_1|$ is usually much smaller than $|V_1|$. In the traveling-wave basis the Hamiltonian is
\begin{equation}\label{eq:H2tilde}
h_1 = \left(\begin{array}{cc}
V_1+U_1 & (V_1-U_1)e^{-i2m\beta_1} \\
(V_1-U_1)e^{i2m\beta_1} & V_1+U_1\\
\end{array}\right) \ ,
\end{equation}
where $\beta_1$ is the azimuthal position of the nanoparticle. With the same procedure for the second nanoparticle and assuming that the coupling between the nanoparticles is negligible we get $H_1 = h_1+h_2$ as in Eq.~(\ref{eq:isolatedcavity}) with
\begin{eqnarray}
\label{eq:defomega}
\Omega_0 & = & \omega_0+V_1+U_1+V_2+U_2 \ ,\\
\label{eq:defA}
A_0 & = & (V_1-U_1)e^{-i2m\beta_1}+(V_2-U_2)e^{-i2m\beta_2} \ ,\\
\label{eq:defB}
B_0 & = & (V_1-U_1)e^{i2m\beta_1}+(V_2-U_2)e^{i2m\beta_2} \ .
\end{eqnarray}
All the quantities above can be computed for the single-nanoparticle case using, e.g., the boundary element method (BEM)~\cite{Wiersig02b}, or approximately using the Green's function approach for point scatterers ($U_j = 0$)~\cite{DMS09}. Note that in general $B_0\neq A_0^*$ as $U_j$ and $V_j$ are complex numbers.

For the remaining elements of the two-cavity coupling matrix~(\ref{eq:twocavitycouplingmatrix}) it is sufficient to discuss the coupling of two disks without nanoparticles. In the standing-wave basis we have
\begin{equation}\label{eq:couplingdoubledisk}
\tilde{h}_3 = \left(\begin{array}{cccc}
\tilde{W}_1 & 0           & W_1         & 0          \\
0           & \tilde{W_2} & 0           & W_2        \\
W_1         & 0           & \tilde{W}_1 & 0          \\
0           & W_2         & 0           & \tilde{W}_2\\
\end{array}\right)
\end{equation}
with $W_j\in\C$ and $\tilde{W}_j\in\C$, where entries vanish since only states with the same parity can couple. If we now assume again that the coupling region of two neighboring cavities is sufficiently long compared to the wavelength~$\lambda$, then $\tilde{W}_1 \approx \tilde{W}_2$ and $W_1 \approx W_2$.

In the traveling-wave basis, this becomes \begin{equation}\label{eq:couplingdoubledisk_t}
h_3 = \left(\begin{array}{cccc}
\tilde{W} & 0         & 0         & W        \\
0         & \tilde{W} & W         & 0        \\
0         & W         & \tilde{W} & 0        \\
W         & 0         & 0         & \tilde{W}\\
\end{array}\right)
\end{equation}
 with $W=\frac{W_1+W_2}{2}\in\C$ and $\tilde{W}=\frac{\tilde{W}_1+\tilde{W}_2}{2}\in\C$.
Note that here one has to flip CW and CCW orientation of one of the cavities to ensure that the unperturbed standing wave modes have the correct symmetry with respect to $x\to-x$.

Again assuming that the couplings ($h_1, h_2. h_3$) are independent of each other we get the two-cavity coupling matrix~(\ref{eq:twocavitycouplingmatrix}) with
\begin{equation}
\Omega = \Omega_0+\tilde{W} \ .
\end{equation}

To confirm  this construction we now consider a particular situation with refractive index $n=2=n_j$, $\beta_1 = 0.7220481635$, $\beta_2 = 1.808017664$, $d_1/R = 0.01$, $d_2/R = 0.02$, $r_1/R = 0.041$, $r_2/R = 0.04857$, $m=16$, and TM polarisation. We first use the BEM to determine $\omega_0$, $V_j$, and $U_j$. Plugging this into Eqs.~(\ref{eq:defomega})-(\ref{eq:defB}) we get
\begin{eqnarray}
\label{eq:para1_1}
A_0 & \approx & 0\\
B_0 & = & 1.189076961\cdot 10^{-3}+i1.076331266\cdot 10^{-5} \ .
\end{eqnarray}
Second, for a double disk system with the disk-to-disk distance $\delta/R = 0.4$, we determine $W_j$ and $\tilde{W}_j$. From this we compute $W$, $\tilde{W}$, and finally $\Omega$:
\begin{eqnarray}
W & = &-0.99104\cdot 10^{-3}-i0.877128797\cdot 10^{-5} \\
\label{eq:para1_4}
\Omega & = & 9.878417152-i0.002293874713 \ .
\end{eqnarray}
Note that we have chosen the parameters such that (i) $|B_0| \gg |A_0|$ and (ii) the phase of $B_0$ and $W$ is similar, which maximizes the effect that we address in the next section.

Using the matrix elements~(\ref{eq:para1_1})-(\ref{eq:para1_4}) the eigenvalues of the model Hamiltonian~(\ref{eq:twocavitycouplingmatrix}) are
\begin{eqnarray}
\Omega_1 & = & 9.879887116-i0.002280860492\\
\Omega_2 & = & 9.876947188-i0.002306888934\\
\Omega_3 & = & 9.878421363-i0.002736868389\\
\Omega_4 & = & 9.878412941-i0.001850881037 \ .
\end{eqnarray}

\begin{figure}[t]
\includegraphics[width=\figurewidth]{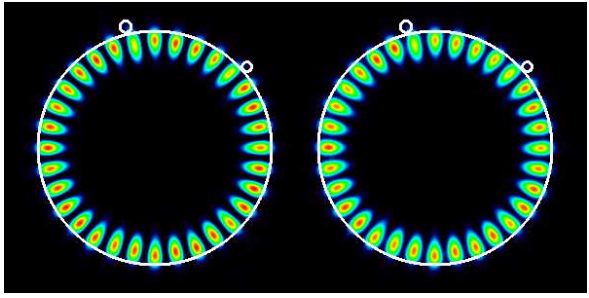}
\caption{(Color online) Intensity $|\psi(x,y)|^2$ of a mode with $m=16$ in a coupled pair of identical microdisks, each perturbed by two nanoparticles.}
\label{fig:mode}
\end{figure}

From the full numerical calculations using the BEM we get for the frequencies of the four nearly degenerate modes
\begin{eqnarray}
\Omega_1 & = & 9.87686556317 -i0.00226714035181\\
\Omega_2 & = & 9.87988254466 -i0.00238812350954\\
\Omega_3 & = & 9.87887740546 -i0.00300042821187\\
\Omega_4 & = & 9.87794940251 -i0.00148975435659 \ .
\end{eqnarray}
Figure~\ref{fig:mode} shows one of the modes. A reasonable agreement between the full numerics and the model Hamiltonian~(\ref{eq:twocavitycouplingmatrix}) can be observed. The deviations are attributed to the four-mode approximation.

\subsection{Band structure}
\label{sec:band}
\begin{figure}[t]
\includegraphics[width=\figurewidth]{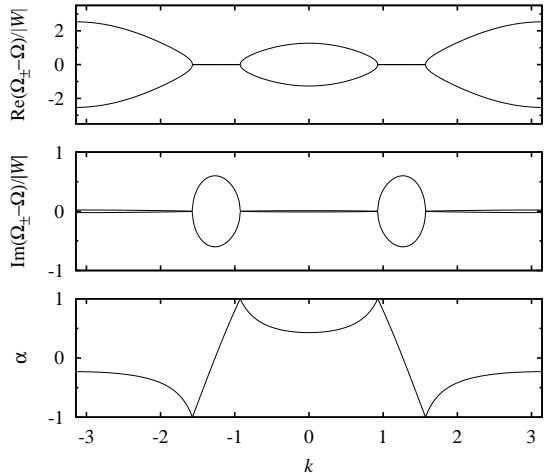}
\caption{The top and middle panel shows the band structure~(\ref{eq:eigenvaluesolution}) for the parameters~(\ref{eq:para1_1})-(\ref{eq:para1_4}). The bottom panel shows the corresponding chirality~(\ref{eq:chirality}).}
\label{fig:CROW1}
\end{figure}

\begin{figure}[t]
\includegraphics[angle=-90,width=\figurewidth]{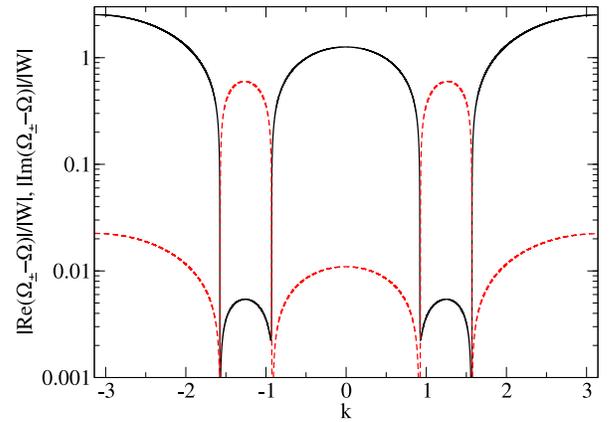}
\caption{Logarithmic representation of the band structure~(\ref{eq:eigenvaluesolution}) for the parameters~(\ref{eq:para1_1})-(\ref{eq:para1_4}); cf. Fig.~\ref{fig:CROW1}. The solid black (red dashed) curve shows the absolute value of the real (imaginary) part of $\Omega_\pm(k)-\Omega$.}
\label{fig:logdCROW}
\end{figure}

Figure~\ref{fig:CROW1}(a) and (b) shows the band structure for the parameters~(\ref{eq:para1_1})-(\ref{eq:para1_4}) in the first Brillouin zone $-\pi\leq k < \pi$. Due to the symmetry with respect to $k=0$ we focus in the discussion on the part with $k\geq 0$. The band structure can be understood in the following way. For the parameter used it holds $A_0\approx 0$ and $B_0\approx -W$. Hence,
\begin{equation}
\Omega_\pm(k) - \Omega \approx \pm \sqrt{2}W\sqrt{\cos{k}}\sqrt{2\cos{k}-1} \ .
\end{equation}
In this case, we find zeroes at $k_1 \approx 1$ and $k_2 \approx 1.57$. As two eigenvalues coalesce, these zeroes are EPs of order $p=2$.
From $|\imag{W}|\ll|\real{W}|$ follows that in the interval $[k_1,k_2]$ the quantity $\Omega_\pm(k) - \Omega$ is nearly purely imaginary which explains the flat band behavior in the top panel of Fig.~\ref{fig:CROW1} in this interval. For the complementary intervals, the quantity $\Omega_\pm(k) - \Omega$ is nearly purely real leading to the flat band behavior in the middle panel of Fig.~\ref{fig:CROW1} in these intervals. Figure~\ref{fig:logdCROW} reveals that the seemingly flat parts in the band structure are in fact slightly curved due to the small but finite size of $\imag{W}$.

\begin{figure}[t]
\includegraphics[angle=-90,width=\figurewidth]{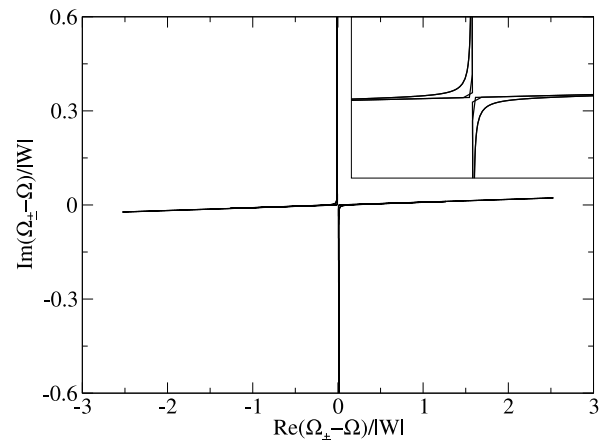}
\caption{Band structure in complex frequency space; cf. Fig.~\ref{fig:CROW1}. The inset shows a magnification of the center region.}
\label{fig:cross}
\end{figure}

Figure~\ref{fig:cross} shows the band structure parametrized by $k\in [-\pi,\pi]$ in the space of complex frequencies. The discussed flat band behavior carries here over to a cross structure with the EPs located in its center. Note that the horizontal line and also the vertical line are slightly tilted due to the small but finite size of $\imag{W}$. This fact implies here that states with larger real part of the frequencies have a lower decay rate.

The  $k$-dependent chirality is shown in the bottom panel of Fig.~\ref{fig:CROW1}. It is remarkable that the chirality is significant for almost all $k$ values. At the EPs the chirality goes to unity, implying that within each resonator the wave travels in a single direction. In the interval $k\in [0,k_1]$ the chirality is positive, i.e., CCW components are larger than CW components. Therefore, in each cavity the light travels mainly in the CCW direction. In the interval $[k_1,k_2]$ the chirality decreases with increasing $k$. The CW components become larger than the CCW components. This predominance prevails in the interval $[k_2,\pi]$.

\subsection{Transport properties}
\label{sec:transport}
\begin{figure}[t]
\includegraphics[width=\figurewidth]{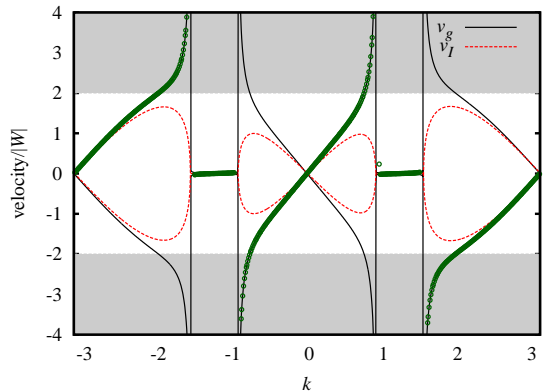}
\caption{(Color online) Group velocity $v_{g\pm}(k)$ (solid black lines) and intensity transport velocity~$v_{I\pm}(k)$ (dashed red lines) for the parameters~(\ref{eq:para1_1})-(\ref{eq:para1_4}). The data points are obtained from the numerical propagation of the wave packet \eqref{eq:initwave} of width $\sigma=80$ in a system of size $N=4000$. The shaded regions denote the forbidden range for $v_I$, cf. Eq.~(\ref{eq:vmax}).}
\label{fig:vgJ}
\end{figure}

Figure~\ref{fig:vgJ} compares the group velocity~$v_{g\pm}(k)$ to the intensity transport velocity $v_{I\pm}(k)$. Both velocities agree at the center and at the border of the Brillouin zone, which is expected from the discussion of Eq.~(\ref{eq:comparison_velocities}). Around the EPs, however, both velocities differ dramatically.
As predicted by Eq.~(\ref{eq:vgEP}) the group velocity diverges at the EP as $(k-k_{\text{EP}})^{-1/2}$. This divergence naturally follows from the singular form of the dispersion relation at an EP and has been also observed numerically in honeycomb photonic lattices~\cite{SRB11}.
The intensity transport velocity always stays finite and at the EP approaches zero  as $(k-k_{\text{EP}})^{1/2}$, which is consistent with Eq.~(\ref{eq:vIEP}). The vanishing of $v_I$ can be related to the full chirality encountered at these points (cf. Fig.~\ref{fig:CROW1}.). It is indeed somewhat surprising that one can have purely CCW (or CW) traveling waves in each cavity having in mind that the coupling between waves of equal sense of rotation is zero (see, e.g., the vanishing entries in the two-cavity matrix~(\ref{eq:twocavitycouplingmatrix})). The interesting conclusion is that at the EP all cavities are effectively decoupled. Nevertheless, as full chirality in each cavity is needed, the cavities are not independent of each other but have to be in the same state. Only in this case, the contributions from adjacent cavities interfere destructively, leading to the effective decoupling.

\begin{figure}[t]
\includegraphics[width=\figurewidth]{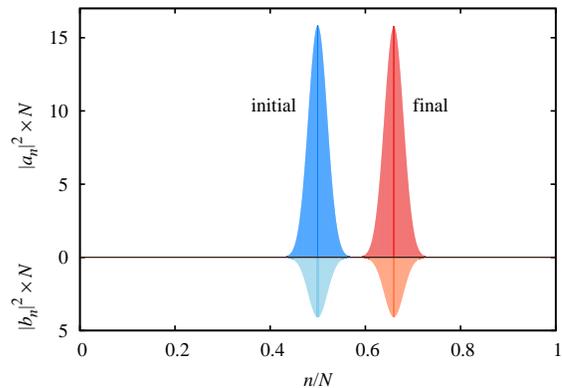}
\caption{(Color online) Initial and final position of the wave packet~\eqref{eq:initwave} of width $\sigma=80$ in a system of size $N=4000$, for $k=0.2\,\pi$ and $A_0=0$. The propagation time is set to $t=8\sigma/|v_g|$.}
\label{fig:wave}
\end{figure}

Under these circumstances, it is interesting to ask which velocity describes the actual motion of a realistic wave packet. To this end, we carried out
numerical calculations of wave packets in a finite system of $N=4000$ resonators. The initial
wave packet is defined in momentum space, according to a discrete-sum approximation of
\begin{equation}
\left(\begin{array}{c}
a_n\\
b_n\\
\end{array}\right)^{\rm init}
= \int dk' \frac{e^{-\sigma^2(k'-k)^2 -i k' (n-N/2)}}{\sqrt{|A(k')|+|B(k')|}}\left(\begin{array}{c}
\sqrt{A(k')}\\
\pm\sqrt{B(k')}\\
\end{array}\right)
\label{eq:initwave}
\end{equation}
with width $\sigma=80$ in lattice-site space. This wave packet is then propagated using the propagation factors $\exp[-i\Omega_\pm(k')t]$ from the dispersion relation (for an illustration see Fig.\ \ref{fig:wave}). The corresponding numerically inferred velocity for one branch of the dispersion is shown as the data points in Fig.~\ref{fig:vgJ}. We find excellent agreement with the group velocity, in agreement with our general considerations in Sec.\ \ref{sec:theory}.

We also considered the wave propagation of an initial wave packet of the form
\begin{equation}
\left(\begin{array}{c}
a_n\\
b_n\\
\end{array}\right)^{\rm init}
= \int dk' \frac{e^{-\sigma^2(k'-k)^2 -i k' (n-N/2)}}{\sqrt{|A(k)|+|B(k)|}}\left(\begin{array}{c}
\sqrt{A(k)}\\
\pm\sqrt{B(k)}\\
\end{array}\right),
\label{eq:initwave2}
\end{equation}
which differs from \eqref{eq:initwave} by the absence of momentum dependence in the Bloch eigenvector in the integral. For the particular parameters chosen here, this wave packet is also found to propagate with $v_g$. However, in the more general cases discussed below such wave packets split up due to the mixing and different losses in the various different bands, which in turn is enhanced due the non-orthogonality of the associated Bloch wave functions.

\begin{figure*}[t]
\includegraphics[width=.95\textwidth]{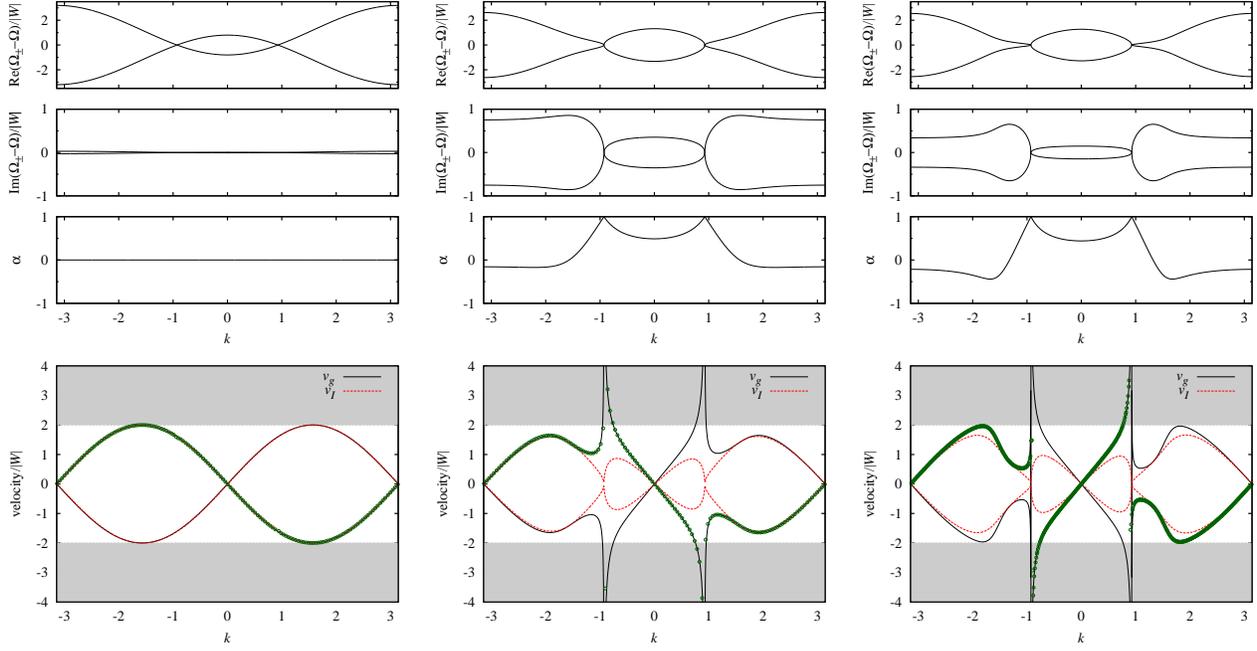}
\caption{(Color online) Band structure and chirality as in Fig.\ \ref{fig:CROW1} and velocities as in Fig.\ \ref{fig:vgJ}, but for the case of fully symmetric backscattering (left column, parameter $A_0$ set to $A_0=B_0$), symmetric internal backscattering (middle column, $A_0=iB_0$)
and in the manifestly asymmetric complex case (right column, $A_0=0.005i \approx iB_0/2$).
\label{fig:CROWsym}}
\end{figure*}

\subsection{Alternative scenarios}
In order to further explore the dependence of the transport properties on the chirality and nonhermiticity we consider three alternative scenarios. In these, the parameters are set to their values in~(\ref{eq:para1_1})-(\ref{eq:para1_4}) with the exception of $A_0$. This parameter is then set either to $A_0=B_0$ (fully symmetric backscattering with $A(k)=B(k)$ throughout the whole Brillouin zone), $A_0=iB_0$ (symmetric internal backscattering with $|A_0|=|B_0|$ for the parameter of the isolated resonators but $|A(k)|\neq |B(k)|$ for the $k$-dependent parameters of the CROW), and $A_0=0.0005i \approx i B_0/2$ (manifestly asymmetric complex case). Note that in all these situations the system is nonhermitian, due to the small imaginary parts in $B_0$ and $W$. The corresponding dispersion relations and velocities are shown in Fig.\ \ref{fig:CROWsym}.
In the totally symmetric case (left column), the band structure remains almost real, and the chirality $\alpha(k)$ vanishes throughout the Brillouin zone. The Bloch vector $\mathbf{v}_\pm=(1,\pm 1)^T/\sqrt{2}$ is $k$-independent, and the velocities $v_g=v_I$ coincide and remain finite. In the internally symmetric case (middle column), the band structure is manifestly complex and displays EPs. The velocities agree well in the regions where the chirality is small, and again drastically differ close to the EPs. The situation is similar in the fully asymmetric complex case shown in the right column.

\section{Concluding remarks}
\label{sec:discussion}
We have derived explicit expressions for the group velocity and the intensity transport velocity of multi-mode wave packets in leaky coupled-resonator optical waveguides. From our analytical results it follows that these two velocities in general differ by a non-hermitian correction to the Hellmann-Feynman theorem. The difference is largest near the exceptional points of the Bloch Hamiltonian. At these points the group velocity diverges, whereas the intensity transport velocity remains finite.

The general theory has been applied to a coupled-resonator optical waveguide made of perturbed microdisks. The perturbation breaks all symmetries and therefore introduces asymmetric internal backscattering, which turns out to be the most relevant source of non-hermiticity. Numerical calculations of the tight-binding model show a complex band structure containing exceptional points where the Bloch modes exhibit a complete chirality. In accordance with the theory, near the exceptional points the group velocity goes to infinity while the intensity transport goes to zero. The latter finding can be related to an effective decoupling of the resonators at the exceptional points. The results are verified with the help of numerical wave packet simulations.

Our general theory  can be directly applied to other coupled-resonator optical waveguides, including to three-dimensional cases such as those made of microspheres. The case of the microsphere is interesting as it is a natural example where more than two modes are present in the unit cell and where higher-order exceptional points can be introduced.
Another class of systems where our theory can be applied is the field of PT-symmetric systems~\cite{BB98}. These are systems with a spectrum that turns  from manifestly real to complex at exceptional points, which then are associated with a form  of spontaneous symmetry breaking. In optical systems this situation can be realized via a balanced arrangement of absorbing and amplifying regions~\cite{EMC07,GSD09,RMG10}. A straightforward application are coupled PT-symmetric dimer chains, which involve a two-mode unit cell~\cite{ZCF10,HS13}. Looking further afield, non-hermitian effects occur in wave transport whenever losses are present. The physics of such systems can be further enriched by nonlinear effects, as they occur, for example, in  chains of coupled quantum dot or quantum well exciton polaritons \cite{Wal13}.

\acknowledgments
We acknowledge financial support by the DFG (grant WI1986/6-1) and EPSRC (grant EP/J019585/1).


\end{document}